# Supervised Heart Rate Tracking using Wrist-Type Photoplethysmographic (PPG) Signals during Physical Exercise without Simultaneous Acceleration Signals

*Mahmoud Essalat, Mahdi Boloursaz Mashhadi,* Student Members, IEEE*, and*
*Farokh Marvasti,* Senior Member, IEEE
Advanced Communications Research Institute (ACRI)
Sharif University of Technology
Tehran, Iran
Email: esalat_mahmoud@ee.sharif.edu

## ABSTRACT

PPG based heart rate (HR) monitoring has recently attracted much attention with the advent of wearable devices such as smart watches and smart bands. However, due to severe motion artifacts (MA) caused by wristband stumbles, PPG based HR monitoring is a challenging problem in scenarios where the subject performs intensive physical exercises. This work proposes a novel approach to the problem based on supervised learning by Neural Network (NN). By simulations on the benchmark datasets [1], we achieve acceptable estimation accuracy and improved run time in comparison with the literature. A major contribution of this work is that it alleviates the need to use simultaneous acceleration signals. The simulation results show that although the proposed method does not process the simultaneous acceleration signals, it still achieves the acceptable Mean Absolute Error (MAE) of 1.39 Beats Per Minute (BPM) on the benchmark data set.

## 1. INTRODUCTION

Heart Rate (HR) is a major vital sign. Accurate HR monitoring is necessary in detection of heart diseases, health monitoring for the elderly, adjusting training load for athletes, and many other applications. With the emergence of wearable sensors and devices such as smartwatches and wristbands, perpetual HR monitoring has become ubiquitous and attracted much attention, recently.

Traditionally HR was estimated using Electrocardiography (ECG) sensors which were attached to the chest and used ground and reference sensors. However, Photoplethysmography (PPG) was later preferred in many application due to its lower cost, easier usage and no requirement of ground and reference sensors [2]. PPG signals can be recorded unobtrusively from different body parts e.g. earlobe, fingertip, and wrist. Hence, the pulse oximeters embedded in smartwatches and wristbands can provide a noninvasive and indirect HR estimation [3].

Unfortunately, PPG signals can be easily contaminated by artifacts caused by hand motions due to the gap between the pulse oximeter and the skin surface. This so called motion artifacts (MA) can also be due to abnormal blood pressure changes in the case that the wearer suffers from blood pressure. Therefore, accurate HR estimation from wrist-type PPG signals especially during intensive physical exercise is challenging.

Recently, Zhang et al [1] proposed a novel method named TROIKA for HR estimation from wrist-type PPG using simultaneous acceleration signals in the fitness scenario in which the subject runs at the maximum speed of 15 km/h and performs deliberate hand swings and sweat whipping from the forehead. TROIKA consisted of Singular Spectrum Analysis (SSA) as a MA reduction step, Sparse Signal Reconstruction (SSR) as a high resolution spectrum estimation stage and Spectral Peak Tracking (SPT) as the last phase for spectral HR corresponding spectral peak finding. Later Zhang et al [4] proposed a new method using JOint Sparse Spectrum estimation (JOSS) in which it has been shown that Multiple Measurement Vector (MMV) can overcome Single Measurement Vector (SMV) used in TROIKA. Both of these methods were computationally expensive which limited their usage in practical works.

The later works enhanced the previous methods both in accuracy and computational performance and reported performance of their proposed methods on the benchmark dataset provided by [1]. Boloursaz et al [5] and Schack et al [6] used multiple adaptive filtering, Zhu et al [7] proposed MICROST which reduced complexity at the cost of increasing the estimation error. Sun et al [8] proposed SPECTRAP based on a novel spectrum subtraction technique using asymmetric least squares and Bayesian decision theory. Although SPECTRAP was accurate and had low computational cost, it could not be used in the causal real time estimation scenario due to the usage of post processing stage. ***CONTRIBUTIONS:*** Many of the previous works have taken heuristic approaches to the problem. This motivated us to look at the problem from a probabilistic point of view, in the sense that it is not possible to claim that a certain frequency corresponds to HR with complete confidence in the strong MA scenarios. However, we can assign a probability to each spectral peak. Hence, in this work we have proposed a novel method based on Neural Network (NN), which can report the confidence score of each HR estimate. The simulation results confirm that the proposed method not only reduces the computational cost, but also achieves an acceptable performance. A major contribution of the proposed technique is that unlike the previous works [1], [4-8], the proposed method alleviates the need to acquire and process simultaneous acceleration signals even during intensive physical exercise. The proposed method still achieves the acceptable MAE of 1.39 BPM on the benchmark data set, without using simultaneous acceleration signals. In addition, our algorithm can automatically recognize the critical frames in which HR estimation is challenging due to intensive MAs and perform a proper combination of the previously proposed MA reduction methods on them. This makes HR estimation very efficient from the run time and accuracy point of view.

The rest of the paper is as follows. In Section 2 we describe the experimental setup, Section 3 introduces the proposed method, Section 4 reports the experimental results and finally Section 5 concludes the paper.

For further reproduction of the reported results, MATLAB codes used in simulations are made available on the authors' personal webpages.

## 2. EXPERIMENTAL SETUP

The data set used in this work is provided by Zhang et al in [1] and was available for Signal Processing Cup 2015. The dataset were collected from 23 subjects, in which first 12 subjects running at the maximum speed of 15 km/h while the rest performed forearm and upper arm exercises e.g. boxing, stretching, shake hands, etc. These subjects carefully worn a wristband embedded with a three-axis accelerometer and two channels of PPG sensors. The signals were collected at the rate of 125 Hz and sent to a nearby computer via Bluetooth. Simultaneously ECG signal was acquired from wet electrodes pasted on the chest of the subjects for extracting ground truth of HR estimation.

## 3. PROPOSED METHOD

### 3.1. Preprocessing

First, the PPG signals are band passed with the lower and upper cutoff frequency of 0.8 and 13 Hz respectively, due to the fact that HR and its first to third harmonic lies in that range. Each PPG signal is divided into 8$s$ signal frames using a sliding window which has a 6$s$ overlap with the previous window and a slide of 2$s$.

### 3.2. Candidate Peak (CP) Selection

In each frame, frequency domain peaks in the normalized $2^{15}$ point autoregressive (AR) spectrum estimation of order 500 of each PPG signal are selected based on the criteria (1):
$$\nabla = CP_1 \cup CP_2$$
$$CP_i = \{\forall\, C_i : M_i > T\} \quad i = 1, 2 \quad (1)$$
In which $C_i$ denotes the frequency location indexes of $i^{th}$ PPG channel, $M_i$ represents the corresponding signal spectrum magnitude, and $CP_i$ is the set of $i^{th}$ PPG channel 'Candidate Peaks', which are indices in spectrum estimation of $i^{th}$ PPG channel that are probable to correspond to HR peak. $T$ is a threshold and has a low amount of 0.3 as HR should have a footprint in the spectrum, however it could be smeared by dominant MA peaks.

Through rest of the paper we use superscript to refer to the member of a set and under script to refer to a specific channel of signal (e.g. $CP_i^k$ is the $k^{th}$ member of $i^{th}$ PPG channel CP and $\nabla^i$ denotes the $i^{th}$ member of $\nabla$). Also, $k^{th}$ harmonic of $CP_i^j$, denoted by $H_i^{jk}$, is defined as follows:
$$H_i^{jk} = argmax_{C_i} M_i :$$
$$C_i \in \left[(k+1) * \left(CP_i^j - \delta\right), (k+1) * (CP_i^j + \delta)\right]$$
Note that a dislocation of $\delta$ is allowed, due to limited signal samples. $\delta$ is set to 18 in our work.

We refer to the spectral magnitude of $H_i^{jk}$ as $S_i^{jk}$.

### 3.3. Feature Extraction

In this stage, the following features will be extracted for each member of $\nabla$. We assume features of $CP_i^j$ are extracting for instance:

*1 & 2-* Magnitude of corresponding $CP_i^j$ frequency location index, denoted by $\widetilde{CP}_i^j$, which we refer to it as $M_i^j$: due to the fact that usually HR related frequency domain peak is great.
$$\widetilde{CP}_i^j = argmax_{C_i} M_i : C_i \in \left[CP_i^j - \delta, CP_i^j + \delta\right]$$

*3 & 4-* Distance between $\widetilde{CP}_i^j$ and $CP_i^j$ denoted by $L_i^j$: they should be close together.
$$L_i^j = \left|\widetilde{CP}_i^j - CP_i^j\right|$$

*5 & 6-* Width and Prominence of the spectral peak corresponding to $CP_i^j$ denoted by $W_i^j$ and $P_i^j$, respectively: maybe HR related frequency domain peak is prominent and narrow.

*7 to 12-* $S_i^{jk} : i = 1,2, k = 1,2,3$ : because Heart Beat (HB) is periodic, which may have strong harmonics.

*13 to 18-* 'Deviation' of $\frac{H_i^{jk}}{k}$ from $CP_i^j$, denoted by $D_i^{jk}$ : according to the fact that they should not have much deviation from each other:

Table 1. Comparison of different features based on achieved J

| Feature | $M_1$ | $M_2$ | $L_1$ | $L_2$ | $W$ | $P$ | $S_1^{j1}$ | $S_1^{j2}$ | $S_1^{j3}$ | $S_2^{j1}$ |
|---|---|---|---|---|---|---|---|---|---|---|
| J | 0.5575 | 0.0507 | 0.0577 | 0.0033 | 0.0093 | 0.1856 | 0.0118 | 0.0109 | 0.0241 | 0.0265 |
| Feature | $S_2^{j2}$ | $S_2^{j3}$ | $D_1^{j1}$ | $D_1^{j2}$ | $D_1^{j3}$ | $D_2^{j1}$ | $D_2^{j2}$ | $D_2^{j3}$ | $N$ | $|\nabla|$ |
| J | 0.0346 | 0.0355 | 0.1450 | 0.0199 | 0.0219 | 0.0555 | 0.0109 | 0.0026 | 0.1455 | 0.1591 |
| Feature | $F^{j1}$ | $F^{j2}$ | $F^{j3}$ | $U$ | $T_1^{j\gamma_1}$ | $T_1^{j\gamma_2}$ | $T_1^{j\gamma_3}$ | $T_2^{j\gamma_1}$ | $T_2^{j\gamma_2}$ | $T_2^{j\gamma_3}$ |
| J | 1.0408 | 1.0000 | 0.9581 | 0.0418 | 0.1335 | 0.1037 | 0.0634 | 0.0928 | 0.0736 | 0.0424 |

$$D_i^{jk} = \left| \frac{H_i^{jk}}{k} - CP_i^j \right|$$

19- $|N_i^j|$ which is the number of 'near' CPs to $CP_i^k$: It is more probable that HR have footprint in both PPG channels compared to MA. This measure can be calculated using $CP_k, i \neq k$. The set of 'Near' CPs to $CP_i^j$, is defined as follows:

$$N_i^j := \{CP_k : |CP_i^j - CP_k| < 24, i \neq k\}$$

20- $|\nabla|$ which is size of $\nabla$: $CP_i^j$ corresponding to PPG spectrums in which $|\nabla|$ is large is less probable to be related to HR.

21 to 23- $F_i^{jk}$ which is the distance between the $CP_i^j$ and the estimated HR in $k^{th}$ previous frame denoted by $HR_{-k}$: Due to large overlap between successive signal frames HR track is smooth. Hence, much distance does not exist between successive HR estimations.

$$F_i^{jk} = \left| CP_i^j - \frac{HR_{-k} * N}{60 * fs} \right|$$

In which $f_s$ is the sampling frequency and $N$ is the number of spectrum estimation points which are 125 Hz and $2^{15}$ respectively in this work.

24- Based on the observation that HR related frequency domain peaks are usually uniquely dominant, 'uniqueness', denoted as $U_i^j$, is defined as:

$$U_i^j := M_i^j - M_i^*,$$
$$M_i^* = \max_k M_i^k : M_i^k < M_i^j, k \neq j, M_i^j, M_i^k \in CP_i$$

25-30- Following features are obtained after normalizing $PPG_i$ signal to its maximum value and using the HR time domain based estimation, denoted as $HR_i^{\gamma_i}$, utilizing number of time domain peaks which have a magnitude greater than $\gamma$ and have a minimum distance of

$$\frac{W_s}{\left(\frac{HR_{-1}*N}{60*f_s} + \tau\right)*f_s} * W_t = \frac{N}{\left(\frac{HR_{-1}*N}{60*f_s} + \tau\right)}$$

In which $W_t$ is the frame length (8s) and $W_s = W_t * f_s$ is the number of frame samples. These features are defined as follows:

$$T_i^{j\gamma_i} = \left| CP_i^j - \frac{HR_i^{\gamma_i}*N}{60*fs} \right|$$

Note that a decrease of $\tau$ in current frequency location index compared to previous HR frequency location index is allowed based on the smooth HR track observation. $\tau$ is set to 30 in our work.

Then we have classified $\nabla$ into HR peak and MA peak classes, assigning 1 to HR peak class and 0 to MA peak class. The assignment is done under criteria (2), using the ground truth as reference and considering the peaks which have a maximum distance of 16 index from corresponding index of true HR as HR peak:

$$\left| \nabla^i - \frac{HR_T * N}{60 * f_s} \right| < 16 \rightarrow HR\ peak \quad (2)$$
$$\left| \nabla^i - \frac{HR_T * N}{60 * f_s} \right| \geq 16 \rightarrow MA\ peak$$

In which $HR_T$ is the true HR in BPM.

### 3.4. Feature Selection

Feature selection is done by the following measure for each feature:

$$J = \frac{(\mu_t - \mu_0)^2 - (\mu_t - \mu_1)^2}{\sigma_0 + \sigma_1}$$

In which $\mu_t$, $\mu_0$ and $\mu_1$ correspond to the mean of corresponding feature on all observations, MA and HR peak classes respectively and $\sigma_0$ and $\sigma_1$ correspond to the variance of MA and HR peak classes respectively. $J$ shows the ability of corresponding feature to separate the two classes.

Table 1 represents $J$ for each feature. The features which have a $J$ greater than 0.3 are selected as final features in our work, resulted in 17 features. It is noteworthy to say that the $21^{th}$ to $23^{th}$ features, however have great $J$ in online processing, will be corrupted by noise if inaccurate estimation is made in previous frames. Hence, they can disturb the whole performance by their huge weights in network, if included. We have exploited their information in Section 3.6.

### 3.5. Neural Network Training

We have used a 3-layer MLP with 22 neurons in its hidden layer [9]. The used NN is as follows:

$$f_{net}^u(w_u, in_u) = \sum_{v \in pred(u)} w_{uv} in_{uv}$$
$$\text{in output layer}: f_{act}^u(net_u, \theta) = net - \theta$$
$$\text{in hidden layer}: f_{act}^u(net_u, \theta) = \tanh(net - \theta)$$
$$f_{out}^u(act_u) = act_u$$

In which $w_{uv}$ and $in_{uv}$ are the weights and inputs to neuron $u$ from its predecessor, neuron $v$, respectively. $f_{net}^u$, $f_{act}^u$ and $f_{out}^u$ are the network input function, activation function and output function of neuron $u$, respectively.

The Results Achieved by the Proposed Algorithm on 23 Subjects' Recordings

| Data No | 01 | 02 | 03 | 04 | 05 | 06 | 07 | 08 | 09 | 10 | 11 | 12 | Mean-12 |
|---|---|---|---|---|---|---|---|---|---|---|---|---|---|
| MAE(Proposed) | 1.76 | 1.44 | 0.79 | 1.32 | 0.88 | 1.21 | 0.80 | 0.75 | 0.67 | 3.81 | 1.16 | 2.08 | 1.39 |
| MAE(TROIKA[1]) | 2.29 | 2.19 | 2.00 | 2.15 | 2.01 | 2.76 | 1.67 | 1.93 | 1.86 | 4.70 | 1.72 | 2.84 | 2.34 |
| MAE(JOSS[4]) | 1.33 | 1.75 | 1.47 | 1.48 | 0.69 | 1.32 | 0.71 | 0.56 | 0.49 | 3.81 | 0.78 | 1.04 | 1.28 |
| MAE(MICROST[7]) | 2.93 | 3.06 | 2.03 | 2.29 | 2.64 | 2.58 | 1.97 | 1.77 | 1.87 | 3.81 | 1.91 | 4.07 | 2.57 |
| ARTPW(Proposed) | 95 | 109.39 | 88.35 | 84.65 | 79.17 | 82 | 71.88 | 79.87 | 77.11 | 104.56 | 83.01 | 101.30 | 88.02 |
| ARTPW(MICROST[7]) | 1.643 | 1.573 | 1.599 | 3.620 | 5.512 | 2.941 | 4.132 | 6.122 | 3.933 | 2.197 | 6.192 | 11.994 | 4.28 |

| Data No | 13 | 14 | 15 | 16 | 17 | 18 | 19 | 20 | 21 | 22 | 23 | Mean-23 |
|---|---|---|---|---|---|---|---|---|---|---|---|---|
| MAE(Proposed) | 11.93 | 2.60 | 2.66 | 12.35 | 5.87 | 1.52 | 3.56 | 3.90 | 1.22 | 0.99 | 1.42 | 2.81 |
| ARTPW(Proposed) | 184.9 | 100.94 | 105.55 | 185.32 | 106.53 | 103.75 | 126.51 | 115.63 | 102.56 | 75.3 | 93.17 | 102.45 |

### 3.6. Framework

In our framework after preprocessing, CP selection in each signal frame, extracting the features of each CP and evaluating them by the trained NN, criteria (3) proposes a technique to detect the signal frames in which further MA reduction is needed to provide a confident HR estimate.

$$If \{\exists\, C^i \in \nabla : O(C^i) \geq 0.4\} \quad (3)$$
$$then\ C^{out} = argmin_{C^i_{up}} \left| C^i_{up} - \frac{HR_{-1}*N}{60*f_s} \right|,$$
$$C^i_{up} \in \{\forall\, C^i : O(C^i) \geq 0.7 * \max_i O(C^i)\}$$

In which $O(C^i)$ is the output of NN for the features extracted from $i^{th}$ member of $\nabla$. $C^{out}$ is the estimated HR corresponding frequency location index.

If criteria (3) is not satisfied then, two enhancement methods are used in succession:
1. Band passing over the following range:
$$\bigcup_{k=1}^{4} [k*f_{-1} - \epsilon, k*f_{-1} + \epsilon]$$

In which $\epsilon$ was set to 0.5 Hz and $f_{-1}$ is the previous window frequency.

This is based on HR track smoothness observation and this method will cause less noise in the features e.g. features *1 & 2* by eliminating MA peaks in HR frequency band.

2. Adding the values of two previous signal frames of a particular PPG channel to its current frame in time domain.

Note that after each method, the algorithm will run again from CP selection stage.

Then based on the smoothness of HR track the algorithm do not permit the estimated HR to be far different from previous HR estimations, based on criteria (4):

$$\left| C^{out} - \frac{HR_{-1}*N}{60*f_s} \right| > \beta \quad (4)$$

In which $\beta$ is set to 75 to limit outliers of NN prediction.

If NN estimation violate criteria (4), then the mentioned enhancement methods will be performed successively, if not performed previously. If two enhancement methods were failed to give an acceptable $C^i$ under the condition (3) then mean of the two previous HR estimation will be reported.

Finally a simple smoothing method is done:

$$HR_{est} = \max(\min\left(\frac{C^{out}*fs*60}{N}, HR_{-1} + \rho\right), HR_{-1} - \rho)$$

In which $HR_{est}$ is the final HR estimation in BPM.

Note that the features $M_1, M_2, L_1, P, D_1^{j_1}, D_2^{j_1}, N, |\Delta|$, and $U$ which are defined independent of the previous HR estimation are used for initializing the algorithm in the first 4 frames.

### 4. SIMULATION RESULTS

To compare the results of our proposed method to the previous works we have used MAE as follow:

$$MAE = \frac{1}{\omega} \sum_{i=1}^{\omega} |HR_{est}(i) - HR_T(i)|$$

In which $(i)$ denotes the $i^{th}$ time window and $\omega$ is the total number of time windows.

It is important to emphasize that although the method proposed in [1] was observed to provide unacceptable results without the SSA stage, in which three-axis accelerometers are used, our method still achieves acceptable results without processing accelerometers.

MAE and the Average Run Time Per Window (ARTPW) in a 8 sec frame is reported for an Intel® Core™ i5-2430M CPU @ 2.40GHz in *ms* in table 2.

### 5. CONCLUSION

In this work we have proposed a novel framework for reliable HR estimation using Neural Network. By simulations on the benchmark datasets, we have demonstrated that the proposed framework achieves acceptable estimation accuracy while keeping low computational complexity. A major contribution of this work is that the proposed method alleviates the need to process simultaneous acceleration signals, even when the user is performing intensive physical exercise. Also unlike other methods, our algorithm can save system resources by recognizing the signal frames in which MA reduction methods are necessary and hence, avoid unnecessary signal processing for the other frames. Furthermore, the proposed method can utilize the previously proposed MA reduction algorithms in the literature, to further improve the performance for these challenging frames.